# Change in the adiabatic invariant in a nonlinear Landau-Zener problem

R. Sokhoyan[1,2,*], D. Melikdzhanian[1], C. Leroy[2], H.-R. Jauslin[2], and A. Ishkhanyan[1]

[1]*Institute for Physical Research of NAS of Armenia, 0203 Ashtarak-2, Armenia*
[2]*Laboratoire Interdisciplinaire Carnot de Bourgogne, UMR 5209 CNRS,
Université de Bourgogne, BP 47870, 21078 Dijon, France*

**Abstract.** We study a nonlinear generalization of the Landau-Zener resonance-crossing problem relevant to coherent photo- and magneto-association of ultracold atoms. Due to the structure of the corresponding classical phase space, the adiabatic theorem breaks down even at very small sweep rates, and the adiabatic approximation diverges because of the crossing of a separatrix. First, by introducing a complex term into the Hamiltonian of the system, we eliminate this divergence and construct a valid zero-order approximation. Further, taking into account that the molecular conversion efficiency and the change of the classical adiabatic invariant at the separatrix crossing are related quantities, we calculate the change of the action for the situation when the system starts from the all-atomic state that corresponds to the case of zero initial action. The absolute error of the presented formula for the change in the action is of the order of or less than $10^{-4}$.

## 1. Introduction

The theory of approximate conservation of adiabatic invariants [1] plays an important role in many domains of physics. According to this theory, the action is an approximately conserved quantity of Hamiltonian systems that contain a slowly varying parameter. This result is based on averaging over the fast motion of a time-independent version of the system (i.e., the same system for which the varying parameter is taken as a constant). The theory states that the change in the action during a time-dependent process is usually on the order of the variation rate of the mentioned slow varying parameter. Moreover, there exists an adiabatic invariant, which is conserved to all orders of the parameter variation rate within time periods no longer than the inverse value of this rate. However, the theory also says that this situation can drastically change if the time-independent version of the system contains separatrices in its phase portrait. In this case the exact phase trajectory of the system may cross the separatrix of the time-independent version of the system. Since the period of the motion along the separatrix is equal to infinity one cannot consider the motion of the time-independent version of the system as a fast one. This results in a breakdown of the averaging method. To describe this case, a rigorous separatrix crossing theory [2-5] has been developed. It has been shown that at the separatrix crossing a jump in the value of the adiabatic invariant occurs, and an asymptotic expression for the value of this jump has been obtained. The separatrix crossing theory has proven to be useful in various problems of plasma physics, hydrodynamics, classical and celestial mechanics, cold molecule formation, etc. (e.g., see Refs. [2-8]).

[*]*Corresponding author: Ruzan Sokhoyan, Institute for Physical Research of NAS of Armenia, 0203 Ashtarak-2, Armenia;*
*phone: +37410533747; fax: +37410585473;*
*e-mail: sruzan@gmail.com*

In the present paper we discuss the coherent formation of ultracold molecules (in particular, molecular condensates) by laser photoassociation [9] or magnetic Feshbach resonance [10]. The production and study of molecular condensates [11] is of vast interest due to important applications, such as ultra-precise molecular spectroscopy and low Doppler width studies of collision processes [12], "superchemical" reactions [13], precision measurements of an electron's electric dipole moment (with certain polar molecules) [14], and quantum computing [15]. Several experimental groups have succeeded in the coherent conversion of a macroscopic number of quantum-degenerate atoms into diatomic molecules starting from either utracold bosonic [16] or degenerate Fermi gases [11]. In Ref. [17] it has been noted that, within the mean-field approximation, *association* of diatomic molecules from degenerate ***Fermi*** gases is mathematically equivalent to *dissociation* of a molecular condensate into ***bosonic*** atoms, and vice versa, *dissociation* of a molecular condensate into degenerate ***Fermi*** atoms is equivalent to *association* of diatomic molecules starting from ultracold ***bosonic*** atoms. The situation we discuss in detail is the coherent association of ultracold bosonic atoms for the case when the external field configuration is defined by the resonance-crossing Landau-Zener model [18].

Long ago, the Landau-Zener model which is well known from the linear theory of nonadiabatic transitions [19] became a standard tool in quantum physics. It describes a situation when two quantum states are coupled by an external field of constant amplitude and a variable frequency, the latter being linearly changed in time. When generalizing the Landau-Zener process to those associated with the mean-field dynamics of interacting many-body systems [20], one obtains nonlinear Landau-Zener processes for which the simple physical intuition based on the linear Landau-Zener model may no longer be valid. The nonlinear version of the Landau-Zener crossing problem has been discussed in numerous papers and elucidated from different points of view (see, e.g., Refs. [21-30]). In particular, it has been shown that, within the framework of the considered model, the change in the action at the resonance passage is a power-law function of the sweep rate through the resonance, as opposed to the exponential law of the linear Landau-Zener problem [18].

Landau-Zener dynamics of diatomic molecule formation from degenerate Fermi-gases has been discussed in Refs. [8,25-30]. In particular, in Ref. [8] the separatrix crossing theory has been employed. One of the main outcomes of Ref. [8] is a formula giving the value of the action jump at the separatrix crossing. This formula contains a parameter referred to as the pseudophase, which strongly depends on the initial conditions. Supposing that the pseudophase is a random variable equally distributed over the open segment $]0,1[$, the authors have succeeded in presenting the dispersion law of the action jump at separatrix crossing.

However, it should be noted that the separatrix crossing theory is not applicable in the case of small initial actions. This kind of situation comes up, e.g., when one considers mean-field dynamics of diatomic molecules formation from ultracold bosonic atoms, if the initial number of molecules is very small or equal to zero. In this case, for the calculation of the action change at a separatrix crossing, a different method has been developed [31] which is based on mapping of the governing equations to a Painlevé equation. Using this method, an asymptotic expression for the action change at the separatrix crossing has been calculated for very slow sweep rates. Interestingly, the mentioned method is also fruitful for the description of diatomic molecules formation from degenerate Fermi-gases [30]; it allows one to generalize and improve the results of Ref. [8].

In the present paper we first discuss the classical phase space of the time-independent version of the problem (for analogous discussions, see, e.g., Refs. [29-30]) and show that the phase trajectory of the time-dependent system will necessarily cross the separatrix of the "frozen" system resulting in the divergence of the adiabatic approximation. Further, we show that it is possible to eliminate this divergence by introducing a complex term into the Hamiltonian of the system, constructing in such a way a valid zero-order approximation. Finally, taking into account that the molecular conversion efficiency is coupled with the change of the action during the whole interaction process we calculate this change in the case when the system starts from the all-atoms state. For arbitrary rates of sweep through the resonance, the absolute error of the presented formula is of the order of or less than $10^{-4}$.

**2. The basic notions and starting equations**

Under the assumption that all the atoms and molecules existing in the system belong to condensates of zero-momentum atoms and molecules, respectively, the coherent conversion of bosonic atoms into diatomic molecules can be described by the following phenomenological momentum-representation two-mode Hamiltonian [32]:

$$\frac{\hat{H}}{\hbar} = \delta_t b_0^+ b_0 + \frac{U}{2\sqrt{N}}(b_0^+ a_0 a_0 + a_0^+ a_0^+ b_0), \qquad (1)$$

where $\hbar$ is Planck's constant, $a_0(a_0^+)$ and $b_0(b_0^+)$ are boson annihilation (creation) operators for zero-momentum atoms and molecules, respectively. The detuning $\delta_t$ defines the energy difference $\hbar\delta_t$ between a zero-momentum molecule and two zero-momentum atoms which can be adjusted by tuning the laser field frequency in the case of photoassociation or by variation of the magnetic field in the case of Feshbach resonance. In the case of photoassociation the atom-molecule coupling $U$ can be controlled by variation of the laser field intensity, while in the case of Feshbach resonance it is a fixed constant. The fact that the Hamiltonian (1) commutes with the operator $\hat{N} = a_0^+ a_0 + 2b_0^+ b_0$, $[\hat{H}, \hat{N}] = 0$, reflects the conservation of the total number of particles $N$, that is, the number of atoms plus twice the number of molecules. A possible way to derive the mean-field equations of motion for the system defined by the Hamiltonian (1) is to write the Heisenberg equations of motion for the operators involved and then replace them by their expectation values. From this approach one can reconstruct the classical Hamiltonian corresponding to the derived set of equations. However, in what follows we apply a different approach: first we define a classical Hamiltonian, corresponding to the second quantization Hamiltonian (1), construct the Poisson brackets for the variables involved and then write the classical equations of motion.

Rescaling the boson operators as $a_0 = a\sqrt{N}$ and $b_0 = b\sqrt{N}$, we rewrite the Hamiltonian in new notations:

$$\frac{\hat{H}}{N\hbar} = \delta_t b^+ b + \frac{U}{2}(b^+ a a + a^+ a^+ b). \qquad (2)$$

It can easily be seen that the rescaled boson operators $a$ and $b$ obey the following commutation relations:

$$[a, a^+] = [b, b^+] = 1/N, \quad [a,b] = [a,b^+] = 0. \qquad (3)$$

For large $N$ ($N \gg 1$) our problem approaches a well-defined classical limit in which the operators can be treated as classical objects. Thus, we replace the operators $a$, $a^+$, $b$, and $b^+$ by $c$-numbers, and the commutators by classical Poisson brackets. This procedure leads to the classical Hamiltonian

$$H = \delta_t b^* b + \frac{U}{2}[b^* a^2 + (a^*)^2 b] \tag{4}$$

and the following expressions for the classical Poisson brackets of the functions $a$ and $b$:

$$\{a, (ia)^*\} = 1, \quad \{b, (ib)^*\} = 1, \quad \{a,b\} = \{a,b^*\} = 0. \tag{5}$$

(the asterisk denotes the complex conjugation). The particle conservation property of the Hamiltonian is now expressed by the following relation:

$$\{H, J\} = 0, \tag{6}$$

where $J = |a|^2 + 2|b|^2$. Equations (5) indicate that the variables $a$ and $(ia)^*$ are canonically conjugate, hence, the Hamiltonian equations of motion in the complex notations are readily written as [33]

$$i\frac{da^*}{dt} = -\partial H/\partial a, \quad \frac{da}{dt} = -i\partial H/\partial a^* \quad \text{and} \quad i\frac{db^*}{dt} = -\partial H/\partial b, \quad \frac{db}{dt} = -i\partial H/\partial b^* \tag{7}$$

($t$ is time). As it can easily be seen, only two of these four equations are independent. Substituting the Hamiltonian (4) into (7), we arrive at the equations of motion within the framework of the mean field approximation:

$$i a_t = U(t) b a^*,$$
$$i b_t = \frac{U(t)}{2} a^2 + \delta_t(t) b. \tag{8}$$

Hereafter the lower-case alphabetical subscript denotes differentiation with respect to corresponding variable. It is convenient to fix the first integral $J$ as $|a|^2 + 2|b|^2 = J = 1$. In this case the function $a$ is interpreted as the atomic state probability amplitude and the function $b$ is, conventionally, interpreted as the molecular state probability amplitude. The quantities $|a|^2$ and $2|b|^2$ are the fractions of atoms and molecules, respectively, with regard to the total number of "atomic particles" $N$ (each molecule is considered as two "atomic particles"). Hence, we refer to $p_1 = |a|^2$ as the atomic state probability and to $p = |b|^2$, conventionally, as the molecular state probability (note that $p_1 \in [0, 1]$ whereas $p \in [0, 1/2]$). We will consider a condensate being initially in all-atomic state: $|a(-\infty)| = 1$, $b(-\infty) = 0$.

The Hamiltonian (4) is defined in a four-dimensional phase space. However, the dimensionality of the phase space can be reduced. To this end, we take into account that $p_1 = 1 - 2p$ and pass to the polar coordinates, thus, representing the probability amplitudes $a$ and $b$ as

$$a = (1 - 2p)^{1/2} e^{i\theta_1}, \quad b = p^{1/2} e^{i\theta_2}, \tag{9}$$

where $\theta_1$ and $\theta_2$ are the corresponding phases. Further, we rewrite the Hamiltonian (4) as follows:

$$H(t,q,p) = \sqrt{p(1-2p)}\, U(t) \cos q + \delta_t(t) p, \qquad (10)$$

where $q = 2\theta_1 - \theta_2$. It can be shown by direct verification that the introduced transformation is canonical, with $q$ and $p$ being the generalized coordinate and the generalized momentum, respectively. Thus, Hamilton's canonical equations,

$$\frac{dq}{dt} = \frac{\partial}{\partial p} H(t,q,p), \quad \frac{dp}{dt} = -\frac{\partial}{\partial q} H(t,q,p), \qquad (11)$$

take the following form:

$$\frac{dq}{dt} = p^{-1/2}\left(\frac{1}{2} - 3p\right) \cdot U(t) \cos q + \delta_t(t),$$
$$\frac{dp}{dt} = p^{1/2}(1-2p) \cdot U(t) \sin q. \qquad (12)$$

In Refs. [29,30] variables analogous to the pair of canonically conjugate variables $\{q, p\}$ has been used for the description of the system's dynamics. However, these variables are not well-defined at $p = 0$ and $p = 1/2$. Since the imposed initial conditions imply that in the beginning of the process $(t \to -\infty)$ should be $p = 0$, it would be more convenient to work with coordinates free of this shortcoming. To this end, we define the following pair of canonically conjugate variables:

$$q' = \frac{1}{\sqrt{2}i}\left(\frac{a}{a^*}b^* - \frac{a^*}{a}b\right), \quad p' = \frac{1}{\sqrt{2}}\left(\frac{a}{a^*}b^* + \frac{a^*}{a}b\right). \qquad (13)$$

Note that the variables $\{q', p'\}$ are related to $\{q, p\}$ as follows:

$$q' = \sqrt{2p}\,\sin q, \quad p' = \sqrt{2p}\,\cos q, \qquad (14)$$

hence,

$$p = \frac{1}{2}\left((p')^2 + (q')^2\right) \quad \sin q = \frac{q'}{\sqrt{(p')^2 + (q')^2}}, \quad \cos q = \frac{p'}{\sqrt{(p')^2 + (q')^2}}. \qquad (15)$$

The variables $\{q', p'\}$ are equal to zero at $p = 0$ but they are not defined at $p = 1/2$ [see Eq. (9)]. In the new coordinates the Hamiltonian is written as

$$H(t,q',p') = \frac{1}{\sqrt{2}} U(t)\, p'[1 - (p')^2 - (q')^2] + \frac{\delta_t(t)}{2}[(p')^2 + (q')^2] \qquad (16)$$

leading to the following equations of motion for the generalized coordinate $q'$ and generalized momentum $p'$:

$$\frac{dq'}{dt} = \frac{1}{\sqrt{2}} U(t)[1 - 3(p')^2 - (q')^2] + \delta_t(t) p',$$
$$\frac{dp'}{dt} = \sqrt{2} U(t) \cdot p'q' - \delta_t(t) q'. \qquad (17)$$

As we see, the new Hamiltonian (16) is a polynomial in terms of $q'$ and $p'$. Note that the function $p'$ satisfies the following nonlinear differential equation of the *second* order:

$$\left(p' - \frac{\delta_t}{\sqrt{2}U}\right)\frac{d^2 p'}{dt^2} + \left(\frac{\delta_{tt}}{\sqrt{2}U} - p'\frac{U_t}{U}\right)\frac{dp'}{dt} - \frac{1}{2}\left(\frac{dp'}{dt}\right)^2 +$$
$$3U^2 \left(p' - \frac{\delta_t - \sqrt{6U^2 + \delta_t^2}}{3\sqrt{2}U}\right)\left(p' - \frac{\delta_t + \sqrt{6U^2 + \delta_t^2}}{3\sqrt{2}U}\right)\left(p' - \frac{\delta_t}{\sqrt{2}U}\right)^2 = 0. \qquad (18)$$

As it is immediately seen, the times for which

$$p' - \frac{\delta_t}{\sqrt{2}U} = 0 \qquad (19)$$

are singular points for this equation. By analyzing the initial set of equations (8) it can be shown that this condition is equivalent to the following one: $2\theta_1 + \delta = C$, where $C$ is a constant.

## 3. The example of the Landau-Zener model and the phase space of the Hamiltonian

As a fundamentally important example of the external field configuration we choose the Landau-Zener model which is defined by

$$U = U_0, \quad \delta_t = 2\delta_0 t. \qquad (20)$$

We first notice that one of the parameters involved in the definition of the model can be eliminated from the equations of motion via rescaling of time. This can be achieved, e.g., by applying the transformation $\tilde{t} = U_0 t$ to the set of equations (17) and introducing the Landau-Zener parameter $\lambda = U_0^2 / \delta_0$. In what follows for simplicity of notations we omit the tilde over $\tilde{t}$. The corresponding Hamiltonian is written as follows:

$$H = \frac{1}{\sqrt{2}} p' \left[1 - (p')^2 - (q')^2\right] + \frac{\gamma}{2}\left[(p')^2 + (q')^2\right], \quad \gamma = \varepsilon t, \text{ and } \varepsilon = 2/\lambda. \qquad (21)$$

From the fact that the Hamiltonian (21) contains only one combined parameter $\lambda$ to characterize the external field one can make an important conclusion: application of high laser field intensities $U_0^2$ along with large sweeping rates $2\delta_0$ or, alternatively, small laser field intensities $U_0^2$ together with small sweep rates $2\delta_0$ will result in the same final molecular population (for $t \to \infty$) provided that the ratio $\lambda = U_0^2/\delta_0$ remains unchanged.

Considering $\gamma$ as a constant, in Fig. 1 we plot the level lines of the Hamiltonian $H = \text{const}$. The level lines represent the phase trajectories of the system for the case when the parameter $\gamma$ is fixed. If the Landau-Zener parameter $\lambda$ is large (i.e. the parameter $\varepsilon$ is small), the parameter $\gamma$ changes slowly in time. In this case we can expect that the system for some time follows a phase trajectory and then slowly passes to another one. Thus, one can imagine that with time the system slowly drifts from one trajectory to another.

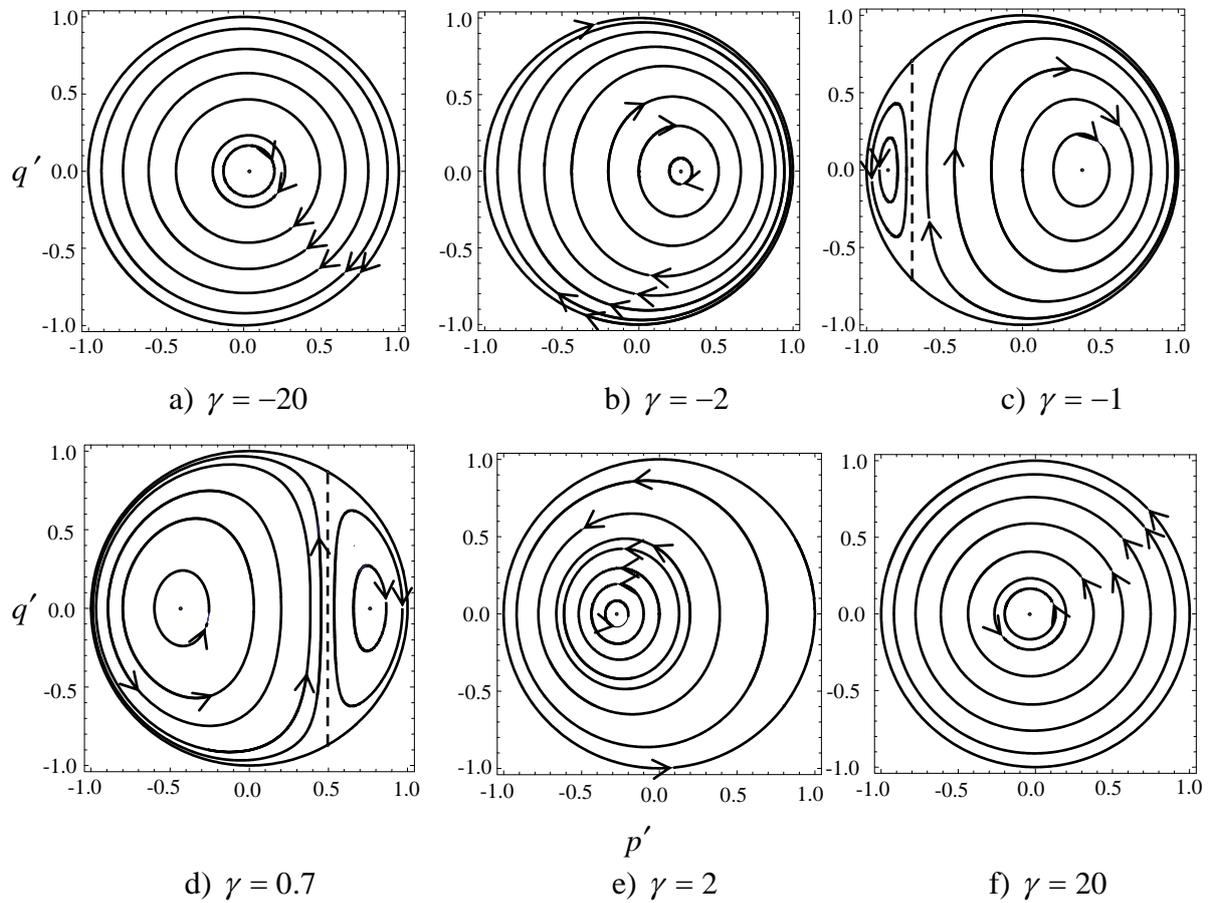

Fig 1. The level lines of the Hamiltonian (21) for different values of the parameter $\gamma$. The dots situated within the phase space represent the two elliptic fixed points (24). The bounding circle of the phase space and the vertical dashed lines of Figs. c) and d) represent the separatrices (27). The vertical dashed line passes through the two saddle fixed points (25) which are situated on the limiting circle of the phase space. The arrows placed along the phase trajectories represent the direction of motion in time.

To understand the structure of the phase space, we analyze the fixed points of the Hamiltonian system under consideration which are defined by equations

$$\frac{\partial H}{\partial q'} = 0, \quad \frac{\partial H}{\partial p'} = 0, \tag{22}$$

that is,

$$\frac{1}{\sqrt{2}}[1 - 3(p')^2 - (q')^2] + \gamma p' = 0, \tag{23}$$

$$q'(\sqrt{2}p' - \gamma) = 0.$$

Solving this set of equations we obtain the following four fixed points:

$$q'_{01,02} = 0, \quad p'_{01,02} = \frac{\gamma \pm \sqrt{6 + \gamma^2}}{3\sqrt{2}}, \tag{24}$$

$$q'_{03,04} = \pm\sqrt{1 - \gamma^2/2}, \quad p'_{03,04} = \gamma/\sqrt{2}. \tag{25}$$

The two points $\{q'_{01}, p'_{01}\}$ and $\{q'_{02}, p'_{02}\}$ are elliptic, while $\{q'_{03}, p'_{03}\}$ and $\{q'_{04}, p'_{04}\}$ are saddle points. The molecular state probability values taken at the considered four fixed points, $p_i = (q'^2_{0i} + p'^2_{0i})/2 \ (i = 1...4)$,

$$p_{1,2} = \frac{1}{18}\left(\gamma^2 + 3 \pm \gamma\sqrt{\gamma^2 + 6}\right) \quad \text{and} \quad p_{3,4} = 1/2, \tag{26}$$

are shown in Fig. 2 as functions of $\gamma$.

Since $p \in [0, 1/2]$, the phase space domain for the variables $p'$ and $q'$ is a disc of radius 1. The bounding circle with a radius of unity corresponds to the all-molecules states, and the center of the circle, $(p', q') = (0,0)$, to the initial all-atoms state (see Fig. 1). For $\gamma \to -\infty$, the fixed point $\{q'_{01}, p'_{01}\}$ is situated in the centre of the phase space circle; it moves to the right with increasing $\gamma$ to stay within the phase space when $\gamma \leq \sqrt{2}$. The fixed point $\{q'_{02}, p'_{02}\}$ appears in the phase space at $\gamma = -\sqrt{2}$ to asymptotically approach the center of the phase space when $\gamma \to +\infty$. As to the saddle points, $\{q'_{03}, p'_{03}\}$ and $\{q'_{04}, p'_{04}\}$, they also appear in the phase space at $\gamma = -\sqrt{2}$ and leave the phase space at $\gamma = \sqrt{2}$. Hence, the points $\gamma = \pm\sqrt{2}$ are bifurcation points of the equations of motion corresponding to the Hamiltonian (21).

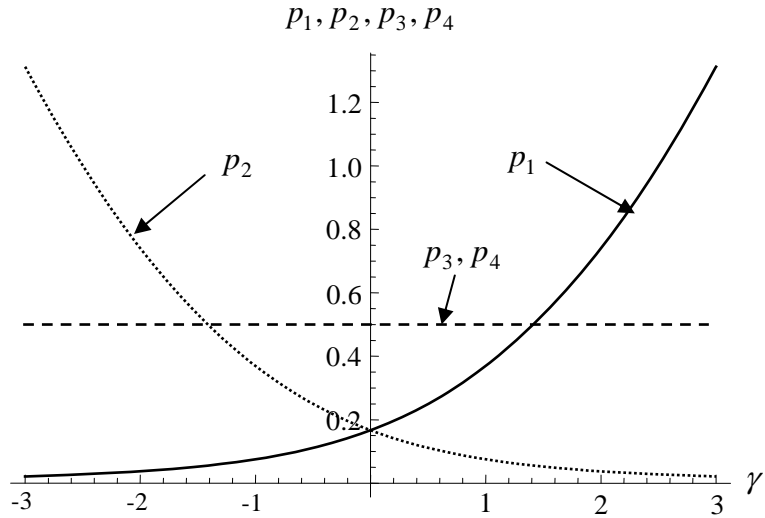

Fig. 2. The values of the transition probabilities (26) at the fixed points (24) $\{q_{1,2}, p_{1,2}\}$ and (25) $\{q_{3,4}, p_{3,4\}}$.

We remark that the coordinate transformation (13) is singular for $a = 0$. However, the set $\{a, b \mid a = 0, |b| = 1/\sqrt{2}\}$ can be identified by continuity with the circle $\{q'^2 + p'^2 = 1\}$, since (13) implies $q'^2 + p'^2 = 2|b|^2$. Therefore, this circle can be incorporated in the definition range of variables $\{q', p'\}$. One can check directly that $\left(a \equiv 0, b = (1/\sqrt{2})e^{i\int \delta_t dt}\right)$ is a solution of the equations of motion (8), for any choice of the time dependent functions $U(t)$ and $\delta_t(t)$. Thus $\{a, b \mid a = 0, |b| = 1/\sqrt{2}\}$ is an invariant set of the dynamics. The circle $\{q'^2 + p'^2 = 1\}$ is in turn invariant with respect to the time evolution of the variables $q'$, $p'$.

The trajectory connecting the two saddle points $\{q'_{03}, p'_{03}\}$ and $\{q'_{04}, p'_{04}\}$ separates the types of motion and is called a separatrix. Taking into account the value of the Hamiltonian at the saddle points, we see that the separatrices correspond to the following trajectories:

$$p' = \gamma/\sqrt{2} \quad \text{and} \quad p'^2 + q'^2 = 1. \tag{27}$$

An important property of the separatrix is that the period along it is equal to infinity. In Figs. 1 c) and d) the separatrices correspond to the vertical dashed line and the phase space limiting circle $p = 1/2$. An interesting observation is that the singular point of the exact equation for $p'$ (18) coincides with the separatrix.

The phase portraits of Fig. 1 describe the dynamics of the system in the case when the external field is defined by the constant-amplitude and constant-detuning Rabi model [36]:

$$U = U_0, \quad \delta_t = 2\delta_0. \tag{28}$$

The nonlinear Rabi problem has been discussed in detail in Ref. [37]. Considering the case when the system starts from the all-atoms state, an exact solution to the problem has been obtained. It has been shown that the molecular state probability $p$ is given in terms of the Jacobi elliptic functions which are periodic in time for arbitrary finite $U_0$ and $\delta_0$, except the case of exact resonance $\delta_0 = 0$. In this case, the Jacobi elliptic function becomes the hyperbolic tangent, and the molecule formation dynamics displays a non-oscillatory behavior approaching the all-molecule state at $t \to +\infty$. By analyzing the phase portraits of the system (see Fig. 1), we can generalize this result to the case of arbitrary initial conditions. Indeed, if the initial conditions are on the separatrix, [i.e., if they satisfy one of the relations (27)] then the exact solution to the problem will not be a periodic function of time, and the molecular state probability will asymptotically go to $1/2$ at $t \to +\infty$.

Finally, we find the exact shape of the phase trajectories at $\gamma \to \pm\infty$ ($\gamma$ is still time-independent). To this end, we consider the asymptotic behavior of the Hamiltonian (21) at $\gamma \to \pm\infty$. The Hamiltonian is then approximated by $H \approx \gamma(p'^2 + q'^2)/2$. Solving Hamilton's equations for this asymptotic Hamiltonian we obtain that for $\gamma \to \pm\infty$

$$p' = \sqrt{2p(\gamma = \pm\infty)}\cos(\gamma t), \quad q' = \sqrt{2p(\gamma = \pm\infty)}\sin(\gamma t). \tag{29}$$

Note that the exact phase trajectories of the system at $t \to \pm\infty$ in the case of the Landau-Zener model ($\gamma = \varepsilon t$) are written as follows:

$$p' = \sqrt{2p(t = \pm\infty)}\cos(\gamma t/2 + \varphi_0), \quad q' = \sqrt{2p(t = \pm\infty)}\sin(\gamma t/2 + \varphi_0), \tag{30}$$

where $\varphi_0$ is an integration constant. Interestingly, the phase trajectories (29) and (30) bound the same area in the phase space (we imply the area inside the phase trajectory circle). Note that in the case of constant $\gamma$ the directions of rotation for $\gamma > 0$ and $\gamma < 0$ are opposite. In the case of the Landau-Zener model, the directions of rotation for $t \to -\infty$ and $t \to +\infty$ are also opposite.

To understand how the nonlinearity affects the dynamics of the system, we analyze the phase space of the linear system

$$\begin{aligned} ia_{L_t} &= U(t)b_L, \\ ib_{L_t} &= U(t)a_L + \delta_t(t)b_L \end{aligned} \tag{31}$$

associated with the nonlinear one under consideration (the question how Eqs. (31) and (8) are related is discussed in detail in Ref. [38]). From the quantum optics point of view this linear system describes coherent interaction of an isolated atom with optical laser radiation [19].

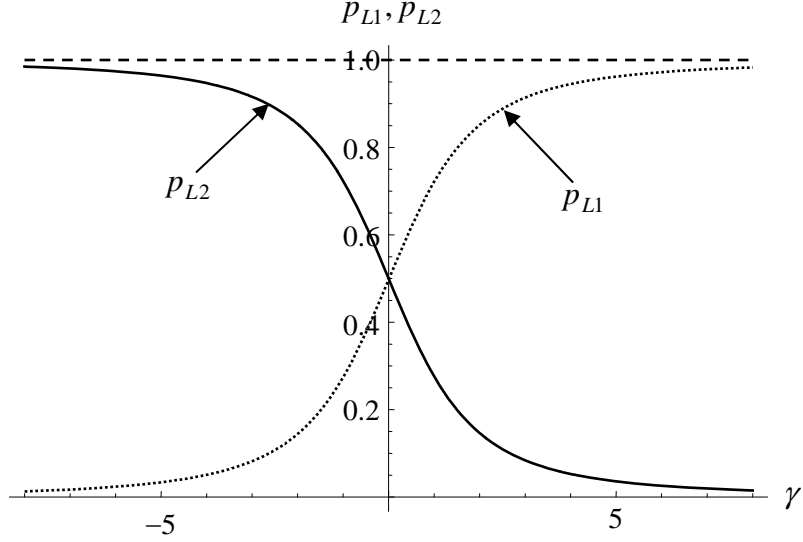

Fig. 3. The values of the transition probability at the fixed points (37) $\{q_{L1},\ p_{L1}\}$ and $\{q_{L2},\ p_{L2}\}$.

This system has the following first integral:

$$|a_L|^2 + |b_L|^2 = \text{const} = J_L \tag{32}$$

which we normalize to unity: $J_L = 1$. Reconstructing the classical Hamiltonian corresponding to the set of equations (31), we arrive at the following result:

$$H_L = \delta_t b_L^* b_L + U[b_L^* a_L + a_L^* b_L] \tag{33}$$

with

$$\{a_L, (ia_L)^*\} = 1, \ \{b_L, (ib_L)^*\} = 1, \ \{a_L, b_L\} = \{a_L, b_L^*\} = 0. \tag{34}$$

In this case, representation of Hamilton's equations of motion in complex notation (7) results in the set of equations (31). Further, representing $a_L$ and $b_L$ as

$$a_L = \sqrt{1-|b_L|^2}\, e^{i\theta_{1L}} \ \text{and}\ b_L = |b_L|\, e^{i\theta_{2L}} \tag{35}$$

we rewrite the Hamiltonian (33) as follows:

$$H_L(t, q_L, p_L) = 2U(t)\sqrt{p_L(1-p_L)}\ \cos q_L + \delta_t(t) p_L, \tag{36}$$

where $p_L = |b_L|^2$ and $q_L = \theta_{1L} - \theta_{2L}$. Further, we write corresponding Hamilton's equations of motion, pass to dimensionless parameters via application of the transformation $\tilde{t} = U_0 t$ to this set of equation, and find the fixed points of the obtained equations. This results in the following two elliptic fixed points:

$$q_{L1,2} = 0, \pi, \quad p_{L1,2} = \frac{1}{2}\left(1 \pm \frac{\gamma}{\sqrt{4+\gamma^2}}\right), \tag{37}$$

where, as before, $\gamma = \varepsilon t$, $\varepsilon = 2/\lambda$, and for simplicity of notations the tilde over $\tilde{t}$ is omitted. The fixed points (37) stay within the phase space for arbitrary values of $\gamma$ (see Fig. 3). Note that the phase space of the time-independent version of the linear system does not contain separatrices. From the set of equations (31) it can be easily seen that in the case of the Rabi model the exact solution to the problem is given in terms of harmonic functions. Hence, non-oscillatory behavior of the second state probability, observed under certain initial conditions in the nonlinear case, is excluded in the linear case.

Now, we again address the nonlinear Landau-Zener problem [see Eqs. (20)-(21)]. Consider that $\gamma$ slowly changes in time. In this case the phase portrait of the system evolves as it is shown in Fig. 1. In the beginning of the process ($t = -\infty$) the phase portrait contains only one elliptic fixed point $\{q'_{01}, p'_{01}\}$ situated in the center of the phase space. At $\gamma = -\sqrt{2}$ bifurcation takes place: another elliptic and two saddle points, $\{q'_{02}, p'_{02}\}$ and $\{q'_{03,04}, p'_{03,04}\}$, respectively, enter the phase space at the point $(q', p') = (0, -1)$. The separatrix moves across the phase space from the left to the right while the elliptic fixed point moves towards the center of the phase space. At $\gamma = \sqrt{2}$ another bifurcation takes place: the elliptic fixed point $\{q'_{01}, p'_{01}\}$ and the two saddle points $\{q'_{03,04}, p'_{03,04}\}$ reach the edge of the phase space and merge at the point $(q', p') = (0, 1)$. When $\gamma > \sqrt{2}$ the phase space again contains only one elliptic fixed point which asymptotically approaches the center of the phase space with $t \to +\infty$. The presented analysis shows that, irrespective of the imposed initial conditions, in the case of the Landau-Zener model the exact phase trajectory will cross the separatrix at a value of $\gamma \in [-\sqrt{2}, \sqrt{2}]$.

## 4. Adiabatic invariance and the nonlinear Landau-Zener problem

In the present section, we construct an approximate solution to the problem using the theory of adiabatic invariants. First, consider the Hamiltonian (21), assuming that $\gamma$ does not vary in time. In this case the action variable, generally defined as

$$I = \frac{1}{2\pi} \oint p' dq', \tag{38}$$

where the integral is taken over a closed trajectory is a conserved quantity of the Hamiltonian system (e.g., see Ref. [1]). Now, suppose that the parameter $\gamma$ slowly varies in time: $\gamma = \varepsilon t$, where $\varepsilon$ is a small parameter. In this case, according to the adiabatic theorem [1], the action (38) is an adiabatic invariant of the system unless the characteristic period of the system is infinity anywhere. Adiabatic invariance implies that for every $\kappa > 0$ there exists $\varepsilon_0(\kappa) > 0$ such that if $0 < \varepsilon < \varepsilon_0$ and $0 < t < 1/\varepsilon$, then

$$|I(t) - I(0)| < \kappa. \tag{39}$$

The proof of the adiabatic theorem is based on an averaging over the fast motion of a time-independent version of the system (i.e., the same system but with $\gamma = \text{const}$). Hence, it is intuitively clear that the parameter $\gamma$ should not change noticeably during a characteristic period of system's motion:

$$T d\gamma / dt \ll \gamma. \tag{40}$$

On the separatrix the characteristic period goes to infinity ($T = \infty$); as a result, when the phase trajectory of the exact system moves across the separatrix of the "frozen" system, the adiabatic theorem is not valid any more, and the action changes its initial value. However, once the system's exact phase trajectory has crossed the separatrix, the adiabatic theorem becomes applicable again and the action again becomes an adiabatic invariant of the system.

If the phase trajectory is closed then the action (38) is nothing else than the $2\pi$ times lesser area of the phase space region bounded by the phase trajectory. But this geometrical definition is not unambiguous. A closed trajectory divides the phase space into two domains and one should specify which domain area should be taken. We define the action as the area such that the domain (whether outer or inner) should always be observed to the right from the path-tracing direction. The action defined in this way will be a continuous function of the parameter $\gamma$. Taking into account that in the case of the Landau-Zener model the exact phase trajectories at $t \to \pm\infty$ are given by Eq. (30), we can easily calculate the exact value of the action $I$ at $t \to \pm\infty$. Indeed, we notice that at $t \to -\infty$ the exact phase trajectory of the system is a circle with an area equal to $2\pi p(t = -\infty)$. Hence, $I(t = -\infty) = p(t = -\infty)$. Further, taking into account the direction of the phase trajectory motion we obtain $I(t = +\infty) = 1/2 - p(t = +\infty)$. Thus, we conclude that the change in the action during the whole interaction process, $I(t = +\infty) - I(t = -\infty)$, can be expressed in terms of the initial ($t = -\infty$) and final ($t = +\infty$) molecular state probabilities:

$$I(t = +\infty) - I(t = -\infty) = 1/2 - p(t = +\infty) - p(t = -\infty). \tag{41}$$

Taking into account the imposed initial condition, $p(t=-\infty)=0$, we have $I(t=-\infty)=0$ and thus

$$I(t=+\infty)=1/2-p(t=+\infty). \tag{42}$$

If in the phase space domains, where the exact phase trajectory is away from the separatrix, the action change is neglected then Eq. (41) can be interpreted as *the action change at the separatrix crossing* written in terms of the initial ($t=-\infty$) and final ($t=+\infty$) molecular state probabilities.

Now, we construct the solution to the problem within the adiabatic approximation and determine its applicability range: this will enable us to find an analytical estimate for the time when the exact phase trajectory crosses the separatrix of the "frozen" system. Recall, that in the beginning of the process ($t=-\infty$) we have $I(t=-\infty)=0$. Since for $\gamma<-\sqrt{2}$ the phase portrait of the "frozen" system does not contain separatrices, the action variable $I$ is an adiabatic invariant of motion, $I(t)=I(t=-\infty)=0$. The action of a trajectory is zero if the trajectory is a fixed point of the system. Hence, in our case the phase trajectory of the system within adiabatic approximation coincides with the first fixed point of Eq. (24). Thus, Eq. (24) defines a solution of the problem within adiabatic approximation while the exact phase trajectory approaches the separatrix.

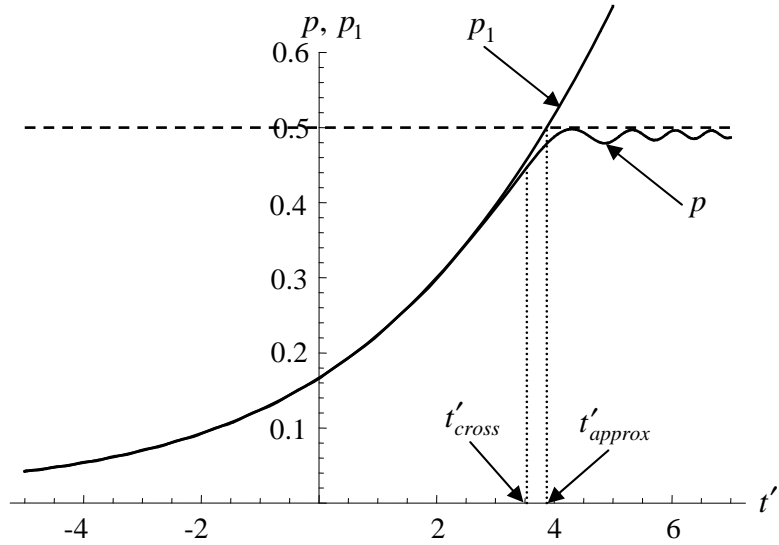

Fig. 4. Numerical graph of the molecular state probability $p$ and adiabatic approximation (26) $p_1$ versus time $t'$ for $\lambda=30$ [where $t'=\sqrt{\delta_0}\,t$ with $t$ being time used in the initial set of equations (8)]. Two vertical lines mark the points where the exact phase trajectory and the adiabatic approximation cross the separatrix, $t'_{cross}\approx 3.535$ and $t'_{approx}=\sqrt{\lambda/2}\approx 3.873$, respectively.

It can be seen that the adiabatic trajectory (24) and the separatrix (27) intersect when $\gamma = \sqrt{2}$. This gives an analytical estimate for the crossing time of the exact phase trajectory and the separatrix of the "frozen" system. In fact, this crossing takes place at a value of $\gamma$ which is smaller than $\sqrt{2}$. This statement is well confirmed by numerical simulations. In Fig. 4 we plot the numerical graph of the molecular state probability $p$ and the adiabatic approximation (26) as functions of time; two vertical lines mark the points where the numerical solution and the adiabatic approximation cross the separatrix of the "frozen" system. As was expected, the adiabatic approximation starts deviating from the numerical solution in the vicinity of the separatrix crossing point. Note that at $\gamma = \sqrt{2}$, $p_1 = 1/2$.

Finally, we apply the presented approach to the linear set of equations (31). In the case when the system starts from the first state, $a_L(t = -\infty) = 1$, $b_L(t = -\infty) = 0$, the phase trajectory of the system in the adiabatic approximation is given by the fixed point $\{q_{L1}, p_{L1}\}$ [see Eq. (37)]. The numerical solutions to the linear problem $p_L$ and the adiabatic approximation (37) are shown in Fig. 5. As we see, in the linear case the application of the adiabatic approximation does not lead to a divergent result. The small-amplitude oscillations emerging after the system has passed through the resonance considerably diminish at larger values of the Landau-Zener parameter $\lambda$; for $\lambda > 3.5$ these oscillations are negligibly small, and the whole temporal dynamics of the system is well described by the adiabatic approximation. In the case of the Landau-Zener model the nonadiabatic corrections for the final transitions probability are exponentially small.

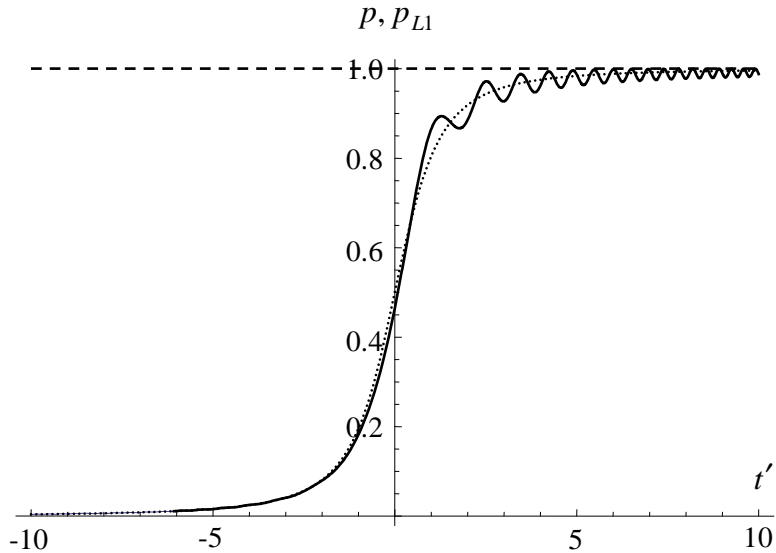

Fig. 5. The numerical solutions to the linear problem (31) $p_L$ (the solid curve) and the adiabatic approximation (37) (the dotted curve) as functions of time $t'$ for $\lambda = 1.7$ [where $t' = \sqrt{\delta_0} t$ with $t$ being time used in the linear set of equations (31)].

## 5. Super-adiabatic sequence and the nonlinear Landau-Zener problem

The adiabatic approximation can be improved by using a general scheme referred to as super-adiabatic approximations [39]. According to this scheme the *n*-th order adiabatic approximation is defined by the recurrence relations

$$\frac{dq'_{n-1}}{dt} = \frac{1}{\sqrt{2}}\left[1 - 3(p'_n)^2 - (q'_n)^2\right] + \gamma\, p'_n,$$

$$\frac{dp'_{n-1}}{dt} = \sqrt{2}\, p'_n q'_n - \gamma\, q'_n. \tag{43}$$

Consider, e.g., the functions $\{p'_{01}, q'_{01}\}$ as a zero-order approximation and, according to the recurrence relations (43), construct the first adiabatic approximation:

$$0 = \frac{1}{\sqrt{2}}[1 - 3(p'_1)^2 - (q'_1)^2] + \gamma\, p'_1,$$

$$\frac{dp'_{01}}{dt} = \sqrt{2}\, p'_1 q'_1 - \gamma\, q'_1. \tag{44}$$

Elimination of $q'_1$ from this system shows that the function $p'_1$ satisfies a polynomial equation of the fourth order:

$$3\left(p'_1 - \frac{\gamma + \sqrt{6+\gamma^2}}{3\sqrt{2}}\right)\left(p'_1 - \frac{\gamma - \sqrt{6+\gamma^2}}{3\sqrt{2}}\right)\left(p'_1 - \gamma/\sqrt{2}\right)^2 + \frac{1}{2}\left(\frac{dp'_{01}}{dt}\right)^2 = 0. \tag{45}$$

Studying now the asymptotic behavior of variables $\{q'_1, p'_1\}$ at $t \to +\infty$ ($\gamma \to +\infty$) for the Landau-Zener model, we see that, at large time values, the roots of equation (45) behave as

$$\left|p'_{1a,1b}\right| \sim \frac{\gamma}{\sqrt{2}}, \quad \left|p'_{1c}\right| \sim \frac{\sqrt{2}}{3}\gamma, \text{ and } \left|p'_{1d}\right| \sim \frac{1}{\sqrt{2}\gamma}, \tag{46}$$

while the function $q'_1$ always tends to zero at $t \to +\infty$ as $|q'_1| \sim C/\gamma$, where $C$ is a constant, depending on choice of the root of (45). Besides, by taking the derivative of Eq. (45) with respect to time it can be shown that there are points of time at which the derivative of $p'_1$ becomes infinite. Finally, as can be seen from Fig. 6, the transition probability in the first adiabatic approximation is not a single-valued function.

As we see, application of the super-adiabatic sequence improves the adiabatic approximation but does not avoid the divergence at the crossing of the separatrix. Hence, alternative approaches are needed. We will show below that introduction of an imaginary term in the Hamiltonian (16) suggests a possibility to construct a zero-order approximation valid for all time.

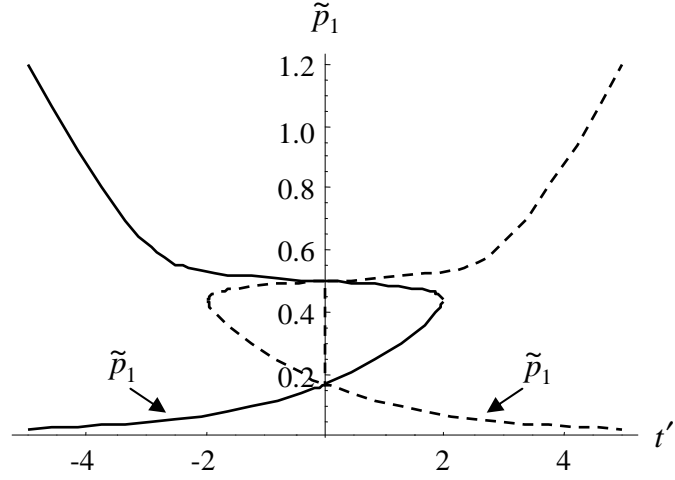

Fig. 6. The transition probability given by the first adiabatic approximation, $\tilde{p}_1 = (q_1'^2 + p_1'^2)/2$, versus time $t'$ ($t' = \sqrt{\delta_0} t$).

## 6. Modification of the adiabatic approximation

In the present section we describe a method which allows us to construct a zero-order approximation valid in the whole time domain. To do this, we analyze the divergence of the adiabatic approximation from the point of view of the theory of ordinary differential equations. From this point of view, when constructing an approximate solution to the set of equations (17), we have neglected the two higher order derivative terms. Thus, the divergence of the approximate solution is due to the singular procedure we have applied to construct it. From this we conclude that, when constructing a zero-order approximation, the higher order derivatives cannot be simply neglected: they should necessarily be taken into account, at least to some extent. We present an approach in which instead of neglecting the derivative $dp'/dt$ in (17), we replace it by some constant $\tilde{A}$, and define a zero-order approximation to the problem under consideration as a function obeying the following set of equations:

$$0 = \frac{1}{\sqrt{2}} U(t)[1 - 3(p_0')^2 - (q_0')^2] + \delta_t(t) p_0',$$
$$\tilde{A} = \sqrt{2} U(t) \cdot p_0' q_0' - \delta_t(t) q_0'.$$
(47)

Studying the solution of these equations, we arrive at an interesting result: the approximate expression for the molecular state probability $\tilde{p}_0 = 1/2[(q_0')^2 + (p_0')^2]$ is a bounded step-like function starting from zero at $t = -\infty$, if and only if $\tilde{A}$ is a pure imaginary constant, i.e., $\mathrm{Re}(\tilde{A}) = 0$ (see Fig. 7). This result is quite unexpected since the quantity $dp'/dt$, which we have tried to approximate by the constant $\tilde{A}$, never takes imaginary values. This statement has been verified for several level crossing models, such as the Landau-Zener and Demkov-Kunike models [40].

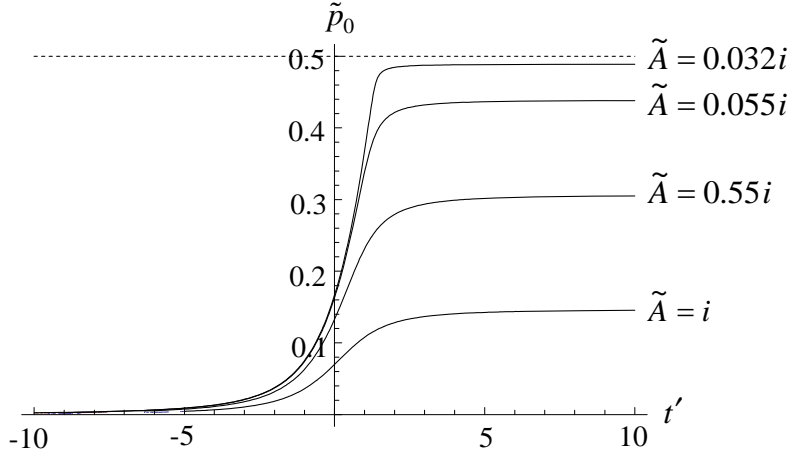

Fig. 7. The approximate expression for the molecular state probability $\tilde{p}_0 = 1/2[(q'_0)^2 + (p'_0)^2]$ in the improved adiabatic approximation (47) versus time $t'$ ($t' = \sqrt{\delta_0}t$), for different values of an imaginary parameter $\tilde{A}$ ($\sqrt{\lambda/2} > |\tilde{A}| > 0$) and fixed $\lambda = 4$.

One can look at the emergence of the constant $\tilde{A}$ from a different angle. Let's consider an effective system described by the following complex Hamiltonian:

$$H(t,q',p') = \frac{U(t)}{\sqrt{2}} p'[1-(p')^2-(q')^2] + \frac{\delta_t(t)}{2}[(p')^2+(q')^2] + \tilde{A}q'. \quad (48)$$

If we now write Hamilton's equations of motion for it and neglect the higher order derivatives, we will again arrive at the set of equations (47). Hence, the valid zero-order approximation (47) has been constructed by introducing a complex term into the Hamiltonian (16).

Note that the precise value of the parameter $\tilde{A}$ has not been specified yet: it plays a role of a variational parameter.

## 7. Review of the results by Ishkhanyan *et al.* [24]

In what follows we juxtapose the presented developments with the results obtained in Refs. [21-24,41], where a completely different approach has been implemented. We show how the results of these works and the presented developments are related and complement each other.

First of all we compare the notations used in the present paper with those of Refs. [24]. Let us apply a unitary transformation to the basic set of equations (8) and write the atomic and molecular probability amplitudes as

$$a = a_1, \quad b = a_2 e^{-i\int \delta_t dt}. \tag{49}$$

This reduces the system (8) to the form used in Refs. [21-24,41]:

$$i\frac{da_1}{dt} = U(t)e^{-i\delta(t)}a_1^* a_2,$$
$$i\frac{da_2}{dt} = \frac{U(t)}{2}e^{i\delta(t)}a_1 a_1, \tag{50}$$

where $\delta$ is the integral of the detuning: $\delta(t) = \int \delta_t dt$. The approach presented in those works is based on the exact equation for the molecular state probability:

$$p_{ttt} - \left(\frac{\delta_{tt}}{\delta_t} + 2\frac{U_t}{U}\right)p_{tt} + \left[\delta_t^2 + 4U^2(1-3p) - \left(\frac{U_t}{U}\right)_t + \frac{U_t}{U}\left(\frac{\delta_{tt}}{\delta_t} + \frac{U_t}{U}\right)\right]p_t +$$
$$+ \frac{U^2}{2}\left(\frac{\delta_{tt}}{\delta_t} - \frac{U_t}{U}\right)(1-8p+12p^2) = 0. \tag{51}$$

When considering the Landau-Zener model (20) in Refs. [21-24] time has been rescaled as $t' = \sqrt{\delta_0} t$ and the dimensionless Landau-Zener parameter $\lambda = U_0^2/\delta_0$ has been introduced. Thus, for the Landau-Zener model, the equation for the molecular state probability (51) takes the following form:

$$p_{ttt} - \frac{p_{tt}}{t} + \left[4t^2 + 4\lambda(1-3p)\right]p_t + \frac{\lambda}{t}(1-8p+12p^2) = 0, \tag{52}$$

where the prime of $t'$ has been omitted.

In Ref. [24], a highly accurate approximate solution of Eq. (52) has been constructed; this approximation is written as a sum of two terms:

$$p = p_0(A,t) + C_1 \frac{p_{LZ}(\lambda_1,t)}{p_{LZ}(\lambda_1,\infty)}. \tag{53}$$

The first term, $p_0(A,t)$, is defined as a solution of an augmented limit equation :

$$\left[4t^2 + 4\lambda(1-3p_0)\right]p_{0t} + \frac{\lambda}{t}(1-8p_0+12p_0^2) - \frac{A}{t} = 0, \tag{54}$$

such that $p_0(A, t=-\infty) = 0$ while the second term, $p_{LZ}(\lambda_1,t)$, is the solution of the *linear*

Landau-Zener problem [18] for an effective Landau-Zener parameter $\lambda_1$:

$$\left(\frac{d}{dt} - \frac{1}{t}\right)\left(\frac{d^2 p_{LZ}}{dt^2} + 4\lambda_1 p_{LZ} - 2\lambda_1\right) + 4t^2 \frac{dp_{LZ}}{dt} = 0. \tag{55}$$

The solution $p_{LZ}$ should satisfy the following initial conditions:

$$p_{LZ}(t)\big|_{t=-\infty} = 0, \quad \frac{dp_{LZ}}{dt}\bigg|_{t=-\infty} = 0. \tag{56}$$

In the case of the Landau-Zener problem, the initial conditions imposed on the molecular state probability and its first derivative [see Eq. (56)] unambiguously define the solution of equation (55). The function $p_{LZ}(\lambda_1, t)$ can be written explicitly in terms of the confluent hypergeometric functions:

$$p_{LZ}(t) = |C_{01} \cdot F_1 + C_{02} \cdot F_2|^2 \tag{57}$$

with

$$C_{01} = \sqrt{\lambda_1 e^{-\pi\lambda_1/4} \cosh(\pi\lambda_1/4)} \, \frac{i}{2} \frac{\Gamma(1/2 - i\lambda_1/4)}{\Gamma(1 - i\lambda_1/4)}, \quad C_{02} = \sqrt{\lambda_1 e^{-\pi\lambda_1/4} \cosh(\pi\lambda_1/4)} \sqrt{i}, \tag{58}$$

and

$$F_1 = {}_1F_1(i\lambda_1/4; 1/2; it^2), \quad F_2 = t \, {}_1F_1(1/2 + i\lambda_1/4; 3/2; it^2), \tag{59}$$

where $\Gamma$ is the Euler gamma-function [35] and ${}_1F_1$ is the Kummer confluent hypergeometric function [35]. The limits of $p_{LZ}$ for $t \to 0$ and $t \to +\infty$ are written as

$$p_{LZ}(0) = \frac{1 - e^{-\pi\lambda_1/2}}{2}, \quad p_{LZ}(+\infty) = 1 - e^{-\pi\lambda_1}. \tag{60}$$

Regarding the limit solution $p_0(A, t)$, integration of Eq. (54) via transformation of the independent variable followed by interchange of dependent and independent variables results in a *quartic* polynomial equation for $p_0$:

$$\frac{\lambda}{4t^2} = \frac{C_0 + p_0(p_0 - \beta_1)(p_0 - \beta_2)}{9(p_0 - \alpha_1)^2(p_0 - \alpha_2)^2}, \tag{61}$$

where $C_0$ is an integration constant and the involved parameters $\alpha_{1,2}, \beta_{1,2}$ are defined as

$$\alpha_{1,2} = \frac{1}{3} \mp \frac{1}{6}\sqrt{1 + \frac{6A}{\lambda}}, \quad \beta_{1,2} = \frac{1}{2} \mp \sqrt{\frac{A}{2\lambda}}. \tag{62}$$

For the initial condition $p_0(-\infty) = 0$ it holds $C_0 = 0$. For a positive $A$, such that $\lambda/2 > A > 0$, the solution of the equation (61) defines a bounded, monotonically increasing function which tends to a finite value less than $1/2$ when $t \to +\infty$ (Fig. 7). In Ref. [24] it has been shown that

$$p_0(+\infty) = \beta_1. \tag{63}$$

By combining Eqs. (53), (62), and (63), it is readily seen that the approximate expression for the final transition probability can be written as follows:

$$p(t = +\infty) = \frac{1}{2} - \sqrt{\frac{A}{2\lambda}} + C_1. \tag{64}$$

In Ref. [24] the following analytic expressions for the variational parameters $A(\lambda)$, $C_1(\lambda)$, and $\lambda_1(\lambda)$ have been obtained:

$$A = \frac{\lambda}{2} {}_2F_1\left(1,2;1.385;-\frac{\lambda^2}{2}\right), \tag{65}$$

$$C_1 = \frac{P_{LZ}(\lambda,+\infty)}{4}\sqrt{{}_2F_1\left(1,2;1.2767;-\frac{\lambda^2}{2.75}\right)}, \tag{66}$$

and

$$\lambda_1 = \lambda\,(1 - 3\beta_1 - 3C_1) = \lambda\,(1 - 3p(+\infty)), \tag{67}$$

where ${}_2F_1$ is the Gauss hypergeometric function [35]. We would like to note that because of the misprint, the third parameter of the hypergeometric function ${}_2F_1$ in the expression for $C_1$ (66) differs from that presented in the original paper [24]. For all the variation range of the input Landau-Zener parameter $\lambda$, the deviation of the formulae (65)-(67) from the numerical results is of the order of or less than $10^{-4}$. Moreover, it has been shown that the absolute error of the analytical formula for the final transition probability with the fitting parameters $A$, $C_1$, and $\lambda_1$ defined by Eqs. (65)-(67) is also of the order of or less than $10^{-4}$.

## 8. Interrelation between the two approaches

Note that if we now take $A = 0$, Eq. (61) will degenerate to a quadratic one because in this case three of four parameters $\alpha_{1,2}$, $\beta_{1,2}$ become equal, $\alpha_2 = \beta_1 = \beta_2 = 1/2$. An interesting observation is that the solutions of this quadratic equation identically coincide with the function (26) derived within adiabatic approximation. Thus, application of the adiabatic approximation is equivalent to removing the two higher order derivative terms from the exact equation for the molecular state probability (52). In this connection it would be interesting to find out whether there exists an analogous interrelation between the function $p_0(A,t)$ [see Eq. (61)] and the improved zero-order approximation for the molecular state probability $\tilde{p} = 1/2[(q'_0)^2 + (p'_0)^2]$, where the functions $\{q'_0, p'_0\}$ are defined as a solution of the set (47) with the initial condition $\tilde{p}(-\infty) = 0$.

To clarify this issue, we rewrite the complex Hamiltonian (48) in terms of the coordinates $\{q, p\}$ [see Eqs. (14)-(15)] and derive the corresponding equations of motion. This results in the following set of equations for the variables $q$ and $p$:

$$\frac{dq}{dt} = p^{-1/2}\left(\frac{1}{2} - 3p\right) \cdot U(t)\cos q + \delta_t(t) + \frac{\tilde{A}}{\sqrt{2p}}\sin q,$$
$$\frac{dp}{dt} = p^{1/2}(1 - 2p) \cdot U(t)\sin q - \tilde{A}\sqrt{2p}\sin q. \quad (68)$$

To construct an improved adiabatic approximation, equivalent to that given by the set of equations (47), we neglect the derivatives of $q$ and $p$ in the system (68). To compare the obtained set of equations with the solution of the augmented limit equation (61), we eliminate the coordinate $q$ from this set. This immediately yields an equation for the determination of the molecular state probability. In the case of the Landau-Zener model the improved adiabatic approximation is written as follows:

$$\frac{U^2(t)}{\delta_t^2(t)} = \frac{p_0(p_0 - \tilde{\beta}_1)(p_0 - \tilde{\beta}_2)}{9(p_0 - \tilde{\alpha}_1)^2(p_0 - \tilde{\alpha}_2)^2}, \quad (69)$$

where

$$\tilde{\alpha}_{1,2} = \frac{1}{3} \mp \frac{1}{6}\sqrt{1 - \frac{6\tilde{A}^2}{\lambda}}, \quad \tilde{\beta}_{1,2} = \frac{1}{2} \mp \frac{i\tilde{A}}{\sqrt{2\lambda}}. \quad (70)$$

If we now choose $\tilde{A}$ as $\tilde{A} = \pm i\sqrt{A}$, the molecular state probability calculated in the improved adiabatic approximation will identically coincide with the solution of the augmented limit equation $p_0(A,t)$. If the negative sign is chosen then $\tilde{\beta}_{1,2} = \beta_{1,2}$; otherwise $\tilde{\beta}_{1,2} = \beta_{2,1}$.

Further, taking into account the fact that the change in the action is coupled with the initial (at $t = -\infty$) and final (at $t = -\infty$) probabilities of the molecular state [see Eqs. (41)-(42)], we use formulae (64)-(67) to obtain the following asymptotic expression for the action variable at $t \to +\infty$.

$$I(t = +\infty) = \frac{1}{2}\sqrt{{}_2F_1\left(1,2;1.385;-\frac{\lambda^2}{2}\right)} - \frac{P_{LZ}(\lambda,+\infty)}{4}\sqrt{{}_2F_1\left(1,2;1.2767;-\frac{\lambda^2}{2.75}\right)}. \quad (71)$$

The absolute error of this formula does not exceed $10^{-4}$, and it is applicable without any limitations on the value of the Landau-Zener parameter $\lambda$. Note that this formula is explicitly expressed in terms of the input parameters of the problem. To write Eq. (71) in terms of elementary functions, we apply the asymptotic expansion for $\lambda \gg 1$ to each of the hypergeometric functions in (71), thus reducing it to the following form:

$$I(t = +\infty) \approx \frac{0.22067}{\lambda} + \frac{1}{\lambda^3}(0.10589 \ln \lambda - 0.24181). \tag{72}$$

The asymptotically exact expression for the action variable at $t \to +\infty$, $I(t = +\infty)$, for very slow resonance sweep rates has been derived in Ref. [31]:

$$I(t = +\infty) = \frac{\ln 2}{\pi} \frac{1}{\lambda} \approx \frac{0.220636}{\lambda} \tag{73}$$

(recall that $\lambda$ is inversely proportional to the resonance sweep rate). As it has been mentioned in Ref. [31], for $\lambda = 20$ numerical calculations reproduce the coefficient $\ln 2 / \pi \approx 0.220636$ with 5-digit accuracy; for larger values of $\lambda$, the absolute error of formula (73) will be even smaller. Thus, as compared to formulas (71) and (72), formula (73) is more precise in the case of very slow sweep rates, but both formulas (71) and even (72) have wider applicability range. For example, the analysis of Eq. (73) indicates that, within the applicability range of the presented formula, the total change in the action, $I(t = +\infty) - I(t = -\infty)$, in the case of large values of the Landau-Zener parameter ($\lambda \gg 1$) can be written as a power-law function of the sweep rate through the resonance. However, Eqs. (71) and (72) clearly show that in the case of arbitrary values of the Landau-Zener parameter the total change in the action is not given by a power-law function of the sweep rate. Moreover, the total change of the action is not given by a power law function of the sweep rate also in the case of the strong interaction limit which corresponds to $\lambda > 1$ (for a detailed discussion of the strong and weak interaction limits see Ref. [24]).

Finally, we conclude this section with some qualitative observations. The exact equation for the molecular state probability (51) indicates that the passage of the system through the resonance can increase the number of equation's singularities: if $\delta_{tt} \neq 0$ at the resonance crossing then the logarithmic derivative of the detuning $\delta_{tt}/\delta_t$ necessarily becomes infinite at this point. Hence, the resonance crossing strongly affects the dynamics of molecule formation. However, the phase space of the time-independent version of the system does not provide a straightforward evidence for the relevance of the resonance crossing. Instead, when studying the phase space of the system, we arrived at a conclusion that the crossing point of the system's exact phase trajectory with the separatrix of the "frozen" system is the essential concept. Indeed, at the separatrix crossing the action substantially changes its value, and due to this crossing the system changes the type of motion. The separatrix crossing point is the singular point of the exact equation for $p'$ (18). A natural conclusion is that the separatrix crossing plays an important role in emergence of the small-amplitude oscillations in the molecular state probability which come up after the separatrix crossing (see Fig. 4).

While studying the linear set of equations (31) we have shown that the corresponding classical phase space does not contain separatrices. However, as it is in the nonlinear case, in the linear case also, certain time after the system has crossed the resonance, small-amplitude oscillations in the molecular state probability appear. In the parameter variation domain, where the adiabatic approximation is applicable (approximately $\lambda > 3.5$), the amplitude of these oscillations is negligibly small, and the function $p_L(t)$ can be regarded as an almost monotonic one. But if we consider the values of the Landau-Zener parameter such that the adiabatic approximation is not applicable (approximately $\lambda < 3.5$), the small amplitude oscillations cannot be neglected (see Fig. 5). By comparing this situation with the one we have in the nonlinear case we see that in the nonlinear case the small-amplitude oscillations cannot be neglected for any values of the Landau-Zener parameter $\lambda$. Hence, we arrive at a conclusion that in the nonlinear case these small-amplitude oscillations are more persistent.

## 9. Conclusion

We have studied the nonlinear mean-field dynamics of molecule formation at coherent photo- and magneto-association of an atomic Bose-Einstein condensate focusing on the case when the external field configuration is defined by the constant-coupling linear resonance-crossing Landau-Zener model. We have studied a condensate initially being in all-atomic state since under contemporary experimental conditions one faces this case most frequently.

Assuming that the sweeping rate through the resonance is small, we have applied the theory of adiabatic invariants. First, we have discussed the classical phase space of the time-independent version of the problem in terms of the canonically conjugate variables $\{q', p'\}$ [see Eq. (13)]. Taking into account that the considered initial condition corresponds to the case of zero initial action we have constructed an expression for the molecular state probability within the adiabatic approximation [see Eq. (26)]. The constructed solution quite accurately describes the temporal dynamics of the coupled atom-molecular system up to the point of time where the approximation, deviating from the numerical solution, starts to go to infinity. Thus, the adiabatic approximation fails to provide a prediction for the final number of the formed molecules.

The reason for the divergence of the adiabatic approximation is that the exact phase trajectory of the system inevitably crosses the separatrix of the system's time-independent version. Hence, the necessary conditions of the adiabatic theorem are not satisfied in this case. However, we have managed to construct a valid zero-order approximation by introducing an imaginary term in the Hamiltonian, writing equations of motion for this augmented Hamiltonian and neglecting the higher order derivative terms. This procedure results in a step-like bounded function that starts from zero. Thus, the introduced complex term has enabled us to eliminate the divergence of the adiabatic approximation.

Further, we have compared the developments of the present paper with those presented in Ref. [24]. We have shown that the application of the adiabatic approximation is equivalent to removing the two higher order derivative terms from the exact equation for the molecular state probability (52) while the constructed zero-order approximation is identical with the solution of the augmented limit equation (54). Taking into account that the molecular conversion efficiency is coupled with the total change of the action $[I(t=+\infty)-I(t=-\infty)]$, we have calculated this change [see Eq. (71)] using a highly accurate approximate formula for the final transition probability presented in Ref. [24]. The absolute error of the presented formula for the action change is on the order of or less than $10^{-4}$. Interestingly, the total change of the action is not given as a power-law function of the sweep rate through the resonance.

Finally, recall that the Hamiltonian we have studied is not restricted to the description of the coupled dynamics of the atomic and molecular condensates only. As it has been mentioned above, it can be mapped to the Hamiltonian describing the formation of ultracold molecules at magneto-association in degenerate Fermi gases [17]. Moreover, it is shown to be equivalent to the time-dependent Dicke model [42]. (Detailed discussion on the correspondence of various quantum models is presented in Ref. [29]). Thus, the results of this paper are equally applicable to all these cases.

**Acknowledgments**

This research has been conducted within the scope of the International Associated Laboratory IRMAS. A. Ishkhanyan acknowledges the support from the Armenian National

Science and Education Fund (ANSEF Grant No. 2009-PS-1692) and the International Science and Technology Center (ISTC Grant No. A-1241). R. Sokhoyan acknowledges the support from INTAS (Young Scientist Fellowship Ref. No. 06-1000014-6484) and the French Embassy in Armenia (Grant No. 2006-4638 Boursière du Gouvernement Français). A. Ishkhanyan acknowledges Institut Carnot de l'Université de Bourgogne for the invited professorship in 2009.